\documentstyle[prl,aps,mypsfig2,twocolumn,mycite]{revtex}

\def\mywidth{\global\columnwidth25.5pc
        \global\hsize\columnwidth\global\linewidth\columnwidth
        \global\displaywidth\columnwidth}

\begin{document}
\mywidth
\preprint{}

\twocolumn
\noindent \large \bf Experimental realization of a quantum algorithm
\vspace*{1ex}\\  
\noindent \small Isaac L. Chuang$^1$, Lieven M.K. Vandersypen$^2$, Xinlan 
Zhou$^2$, Debbie W. Leung$^3$, {\em and}\/ Seth Lloyd $^4$ \vspace*{0.5ex}\\
\scriptsize $^1$ IBM Almaden Research Center K10/D1, San Jose, CA 95120, 
$^2$ Solid State Electronics Laboratory, Stanford University, 
Stanford, CA 94305 $^3$ Edward L. Ginzton Laboratory, 
Stanford University, Stanford, CA 94305
$^4$ MIT Dept. of Mechanical Engineering, Cambridge, Mass. 02139 
\vspace*{0.3ex}\\
\scriptsize submitted January 19, 1998; revised March 7, 1998, accepted March 18, 1998.

\def\>{\rangle}
\def\be{\begin{equation}}
\def\ee{\end{equation}}
\def\bea{\begin{eqnarray}}
\def\eea{\end{eqnarray}}
\newcommand{\ket}[1]{\mbox{$|#1\rangle$}}
\newcommand{\bra}[1]{\mbox{$\langle #1|$}}
\newcommand{\mypsfig}[2]{\psfig{file=#1,#2}}



\vspace*{-10ex}

\small

{
\bf
A quantum computer is a device that processes information in a
quantum-mechanically coherent fashion
\cite{Deutsch85,Shor94,Divincenzo95,Lloyd95b,Ekert96}. In principle, it can
exploit coherent quantum interference and entanglement to
perform computations, such as factoring large numbers or searching an unsorted
database, more rapidly than classical computers
\cite{Deutsch85,Shor94,DeutschJozsa,Simon94,Grover}. Noise, decoherence, and
manufacturing problems make constructing large-scale quantum computers
difficult\cite{Unruh95,Chuang95a,Landauer88,Landauer95,Palma96}. Ion traps 
and optical cavities offer promising experimental approaches
\cite{Monroe95,Turchette95}, but no quantum algorithm has yet been implemented
with those systems.
On the other hand, because of their natural isolation from the environment,
nuclear spins are particularly good `quantum bits'\cite{Lloyd93}, and their
use for quantum computation is possible by applying nuclear magnetic
resonance (NMR) techniques in an unconventional
manner\cite{Gershenfeld97,Cory97b,Cory97a}.  Here, we report on the
experimental realization of a quantum algorithm using NMR, to solve a purely
mathematical problem in fewer steps than is possible classically.  In
particular, our simple quantum computer can determine global properties of
an unknown function using fewer function `calls' than is possible using a
classical computer.
}

We implemented the simplest possible version of the Deutsch-Jozsa (D-J)
quantum algorithm\cite{DeutschJozsa}, which determines whether an unknown
function is constant or balanced. A constant function $f(x)$ from $N$ bits to
one bit either has output $f(x)=0$ for all $x$, or $f(x)=1$ for all $x$.  A
balanced function has $f(x)=0$ for exactly half of its inputs, and $f(x)=1$
for the remaining inputs.  To determine with certainty whether a function is
constant or balanced on a deterministic classical computer, requires up to
$2^{N-1} + 1$ function calls: even if one has looked at half of the inputs and
found $f(x)=0$ for each, one still can't conclude with certainty that the
function is constant.  In contrast, the D-J algorithm, as improved by
R. Cleve, {et al.}~\cite{Cleve97} and Alain Tapp, allows a quantum computer to
determine whether $f(x)$ is constant or balanced using only one function call.

\begin{figure}[htbp]
\begin{center}
\mbox{\psfig{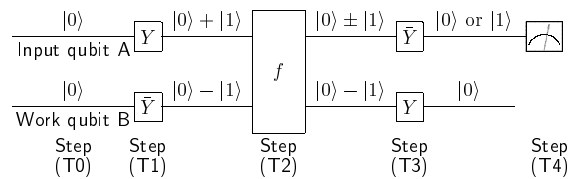}}
\end{center}
\caption{Quantum circuit for performing the D-J algorithm.  {\bf (T0)} Start
 with both the `input' and `work' qubits (A and B) in the state $|0\rangle$.
 {\bf (T1)} Perform the transformation $Y$:$|0\>\rightarrow \big( |0\> +
 |1\>\big)/\sqrt{2},$ $|1\>\rightarrow \big( -|0\> + |1\>\big)/\sqrt{2},$ to A,
 and the inverse transformation $\bar Y$ to B, resulting in the state
 $\frac{1}{2} \sum_{x=0}^{1} |x\> \left( |0\> - |1\> \right)$.  The input qubit
 in some quantum sense registers both 0 and 1 at once.  {\bf (T2)} Call the
 function: apply $f$ to A, and add the result to B modulo 2.  As long as the
 quantum logic operations needed to evaluate $f$ are carried out coherently,
 the work qubit now contains in some quantum sense the outputs of $f$ on all
 possible inputs, an effect that Deutsch termed `quantum
 parallelism'\protect\cite{Deutsch85}. The two qubits are now in the state
 $\frac{1}{2} \sum_{x=0}^{1} |x\> \left( |0+f(x)\> - |1+f(x)\>\right)$ $=
 \frac{1}{2} \sum_{x=0}^{1}(-1)^{f(x)} |x\> \left( |0\> - |1\>\right)$.  {\bf
 (T3)} Perform the inverse of the transformations of (T1), thereby taking the
 qubits out of their superposition states.  If $f$ is constant, then the
 factors $(-1)^{f(x)}$ are either all $+1$ or all $-1$, and the result of the
 transformation in this step is the state $\pm |00\>$.  If $f$ is balanced,
 then exactly half of the factors $(-1)^{f(x)}$ are $+1$ and half are $-1$, and
 the result of the transformation is the state $\pm |10\>$.  {\bf (T4)} Read
 out A.  If it is 0, then $f$ is constant; if it is 1, then $f$ is balanced.}
\label{fig:dj_pulse_outline}
\end{figure}

The D-J algorithm is well illustrated by its simplest possible case, when $f$
is a function from one bit to one bit; this is the version that we have
realized (it is also the simplest instance of Simon's
algorithm\cite{Simon94}).  There are four possible $f$'s, two of which are
constant, $f_1(x)=0, f_2(x)=1$ and two of which have an equal number of 0 and
1 outputs: $f_3(x)=x, f_4(x)= {\tt NOT}~x$.  To determine whether such a
function is constant or balanced is analogous to determining whether a coin is
fair -- with heads on one side and tails on the other; or fake -- with heads
on both sides.  Classically, one must look at the coin twice, first one side
then the other, to determine if it is fair or fake.  The D-J algorithm
exploits quantum coherence to determine if a quantum `coin' is fair or fake
while looking at it only once.  The algorithm requires one `input' spin and
one `work' spin, and is schematically represented by the quantum circuit shown
in Fig.~\ref{fig:dj_pulse_outline}.

Experimentally, this quantum algorithm was implemented using the nuclear spins
of the $^1$H and $^{13}$C atoms in a carbon-13 labeled chloroform molecule
(CHCl$_3$) as the input and work quantum bits (`qubits').  $|0\>$ ($|1\>$)
describes the spin state aligned with (against) an externally applied, strong
static magnetic field ${\bf B}_0$ in the $+ \hat z$ direction.  The reduced
Hamiltonian for this 2-spin system is to an excellent approximation given by
($\hbar=1$)~\cite{Slichter}
\be
	\hat{\cal H} = - \omega_A \hat{I}_{zA} - \omega_B \hat{I}_{zB} +
	2\pi J \hat{I}_{zA} \hat{I}_{zB} + \hat{\cal H}_{env} 
\,.
\label{eq:nmrhamiltonian}
\ee
The first two terms describe the free precession of spin $A$ ($^1$H) and $B$
($^{13}$C) about ${\bf -B}_0$ with frequencies $\omega_A/2\pi$ $\approx$ 500
MHz and $\omega_B/2\pi$ $\approx$ 125 MHz. $\hat I_{zA}$ is the angular
momentum operator in the $+ \hat z$ direction for $A$. The third term
describes a scalar spin-spin coupling of the two spins of $J$ $\approx$ 215
Hz. $\hat{\cal H}_{env}$ represents couplings to the environment, including
interactions with the chlorine nuclei, and also higher order terms in the
spin-spin coupling, which can be disregarded (as will be described below).


The five theoretical steps of the quantum algorithm, (T0)--(T1), were
experimentally implemented as follows:

\noindent {\bf (E0)} An input state is prepared with a 200 mM, 0.5 ml sample
of chloroform dissolved in d6-acetone, at room temperature and standard
pressure.  The ${\cal O}(10^{18})$ molecules in this bulk sample can be
thought of as being independent single quantum computers, all functioning
simultaneously.
The theoretically ideal result is obtained when the spins in all the 
molecules start out in the
$00$ state.  Because the experiment is performed at room temperature, however,
the initial density matrix $\rho$ for the thermally equilibrated system has
populations ${\rm diag}(\rho) = [n_{00},n_{01},n_{10},n_{11}]$ in the $00$,
$01$, $10$, and $11$ states, respectively, where $\rho$ is the density matrix,
and $n_i$ are proportional to $e^{-E_i/kT}/2^N \approx (1-E_i /kT)/2^N$, with
$E_i$ the energy of state $i$ and $N=2$ being the number of qubits used in our
experiment.
A variety of techniques exist to extract from this thermal state just the
signal from the $00$ state\cite{Gershenfeld97,Cory97b}; we adopted the method
of `temporal averaging'\cite{Knill97}, which involves the summation of three
experiments in which the populations of the $01$, $10$, and $11$ states are
cyclically permuted before performing the computation.  The essential
observation is that $[n_{00},n_{01},n_{10},n_{11}] +
[n_{00},n_{11},n_{01},n_{10}] + [n_{00},n_{10},n_{11},n_{01}] = \alpha
[1,1,1,1] + \delta [1,0,0,0]$, where $\alpha = n_{01}+n_{10}+n_{11}$ is a
background signal which is not detected, and $\delta = 3n_{00} - \alpha$ is a
deviation from the uniform background whose signal behaves effectively like
the desired pure quantum state, $|00\>$.  The permutations are performed using
methods similar to those used for the computation, described next.  This
technique avoids the technical difficulties of detecting the signal from a
single nuclear spin, and allows a sample at room temperature, which produces
an easily detectable signal, to be used for quantum computation.

Note that while this method requires $f(x)$ to be evaluated 3 times, it is
actually not necessary.  Although step (T0) stipulates an input pure state
$|00\>$, the algorithm works equally well if the input qubit is initially
$\ket{1}$; furthermore, when the work qubit is initially $\ket{1}$, it fails,
and cannot distinguish constant from balanced functions, but this does not
interfere with other computers which have worked. Thus, a thermal state is a
good input for this algorithm, and only one experiment needs to be performed.
Data from both thermal and pure state inputs are presented below.

\noindent {\bf (E1)} Pulsed radio frequency (RF) electromagnetic fields are
applied to transform the qubits as prescribed in {(T1)}.  These fields,
oriented in the $\hat x-\hat y$ plane perpendicular to ${\bf B}_0$,
selectively address either $A$ or $B$ by oscillating at frequency $\omega_A$
or $\omega_B$.  Classically, an RF pulse along $\hat y$ (for example) rotates
a spin about that axis by an angle proportional to $\approx t P$, the product
of the pulse duration $t$ and pulse power $P$.  In the `bar magnet' picture,
a $\pi/2$ pulse along $\hat y$ (we shall call this $Y$) causes a $\hat z$
oriented spin to be rotated by $90^\circ$, onto $\hat x$ (similarly, we shall
let $\bar Y$ denote $\pi/2$ rotations about $-\hat y$, and $X$ denote $\pi/2$
rotations about $\hat x$, and so forth; subscripts will identify which spin
the operation acts upon).  This description of the state is classical in the
sense that a bar magnet always has a definite direction.  In reality, however,
a nuclear spin is a quantum object, and instead of being aligned along $\hat
x$, it is actually in a superposition of being up and down,
$(|0\>+|1\>)/\sqrt{2}$.  Likewise, a spin classically described as being along
$-\hat x$ is actually in the state $(|0\>-|1\>)/\sqrt{2}$.  {(E1)} thus
consists of applying the two RF pulses $Y_A {\bar Y}_B$.

\noindent {\bf (E2)} The function $y \rightarrow y \oplus f(x)$ is implemented 
using RF pulses and spin-spin interaction.  Recall that spin $A$ represents 
the input qubit $x$, and $B$ the work qubit $y$ where $f$ stores its output.
$f_1$ is implemented as $\tau/2 - X_B X_B - \tau/2 - X_B X_B$, to be read from
left to right, where $\tau/2$ represents a time interval of $1/4J \approx
1.163$ ms, during which coupled spin evolution occurs.  Dashes are for
readability only, and typical pulse lengths were $10$-$15$ $\mu$s.  This is a
well known refocusing\cite{Ernst94} pulse sequence which performs the identity
operation.  $f_2$ is $\tau/2 - X_B X_B - \tau/2$, similar to $f_1$ but without 
the final pulses, so that $B$ is inverted.  $f_3$ is $Y_B - \tau -
\bar Y_B X_B - \bar Y_A \bar X_A Y_A$, which implements a `controlled-{\sc
not}' operation, in which $B$ is inverted if and only if $A$ is in the $|1\>$
state.  The naive `bar magnet' picture can be used to get a feeling for how
this works in case the inputs are $00$ or $10$, for which the 
subsequence $Y_B-\tau-X_B$ suffices (note that after {(E1)}, both spins 
are not just $\ket{0}$ or $\ket{1}$ but in a superposition of both, in which 
case the extra pulses of $f_3$ are necessary~\cite{Gershenfeld97}). First, 
$Y_B$ rotates $B$ to $+\hat x$. $B$ then precesses in the $\hat x
-\hat y$ plane, about $-\hat z$. Due to the spin-spin coupling, $B$ precesses 
slightly slower (faster) if $A=0$ ($A=1$).  After $\tau$ seconds, $B$ reaches
$+\hat y$ ($-\hat y$) in the rotating frame. $X_B$ then rotates $B$ to 
$+\hat z$ ($- \hat z$), i.e. to $0$ or $1$, where the final state of $B$
depends on the input $A$.  The precise quantum
description is easily obtained by multiplying out the unitary rotation
matrices.  Finally, $f_4$ is implemented as $Y_B - \tau - \bar Y_B \bar X_B -
\bar Y_A \bar X_A Y_A$, which is similar to $f_3$ but leaves $B$ inverted.

\noindent {\bf (E3)} The inverse of {(E1)} is done by applying the RF pulses
$\bar{Y}_A Y_B$ to take both spins back to $\pm \hat z$. Spin $A$, which was
$\ket{0}$ at the input, is thus transformed into $\ket{0}$ or $\ket{1}$ for
constant or balanced functions respectively.

\begin{figure}[htbp]
\centerline{\mbox{\psfig{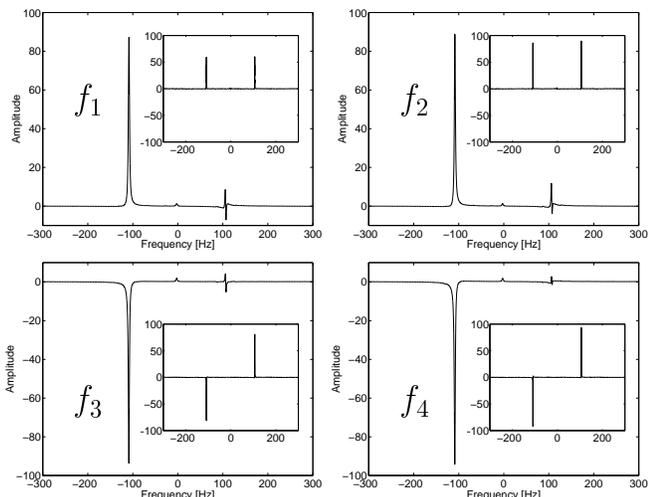}}}
\caption{Proton spectrum after completion of the D-J algorithm and a single
 read-out pulse $X_A$, with an effectively pure input state $\ket{00}$ and with
 a thermal input state [Inset].  The low (high) frequency lines
 correspond to the transitions $\ket{00}\leftrightarrow\ket{10}$
 ($\ket{01}\leftrightarrow\ket{11}$). The frequency is relative to 499755169
 Hz, and the amplitude has arbitrary units.  The spectrum is the Fourier
 transformed time varying voltage $V(t)$, induced in the pick up coil by the
 precession of spin $A$ about ${\bf -B}_0$, at frequency $\omega_A$, after the
 read-out pulse $X_A$.  $V(t)$ is given by $V(t) \approx V_0 \,{\rm Tr} \,
 [e^{-i\hat{\cal H}t}e^{-i\frac{\pi}{2}\hat{I}_x} \rho(0)
 e^{i\frac{\pi}{2}\hat{I}_x}e^{i\hat{\cal H}t} \times (-i \hat\sigma_{xA} -
 \hat\sigma_{yA})]$, where $\hat\sigma_{\{x,y\}}$ are Pauli matrices, and
 $\rho(0)$ is the density matrix of the state immediately before the readout
 pulse.  By this convention, a spectral line for spin $A$ is real and positive
 (negative) when spin $A$ is $|0\>$ ($|1\>$) right before the $X_A$ read-out
 pulse.  Experiments were performed at Stanford University using an 11.7 Tesla
 Oxford Instruments magnet and a Varian $^{\sf UNITY}${\sl Inova} spectrometer
 with a triple-resonance probe. $^{13}$C-labeled CHCl$_3$ was obtained from
 Cambridge Isotope Laboratories, Inc. [CLM-262].}
\label{fig:spectra}
\end{figure}

\noindent {\bf (E4)} The result is read out by applying a read-out pulse $X_A$
to bring spin $A$ back into the $\hat x-\hat y$ plane. The time varying 
voltage $V(t)$ induced by the precession of spin $A$ about ${\bf -B}_0$ is
recorded by a phase sensitive pick-up coil. Inspection of the spectrum of 
$V(t)$ after a single experiment run and an appropriate read-out pulse, 
immediately reveals whether $f(x)$ is constant or balanced, as shown in 
Fig.~\ref{fig:spectra}.

\begin{figure}[t]
\centerline{\mbox{\psfig{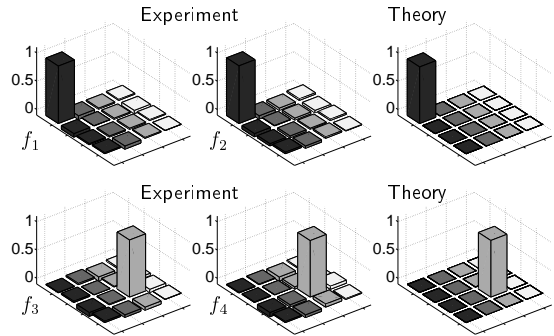}}}
\caption{Experimentally measured and theoretically expected deviation density
	matrices after completion of the D-J algorithm.  The
	diagonal elements represent the normalized populations of the states
	$\ket{00}, \ket{01}, \ket{10}$ and $\ket{11}$ (from left to
	right). The off-diagonal elements represent coherences between
	different states. The magnitudes are shown with the sign of the real
	component; all imaginary components were small. The deviation density 
	matrix was obtained from the integrals of the proton and carbon 
	spectral lines, acquired for a series of 9 experiments with different 
	read-out pulses for each spin (quantum state
	tomography\protect\cite{Chuang97e}). 
 The observed experimental non-idealities can be quantified as follows.  In the
 experiments, the normalized pure-state population (ideally equal to $1$),
 varied from $0.998$ to $1.019$. The other deviation density matrix elements 
 (ideally $0$), were smaller than $0.075$ in magnitude.  The relative error 
 $\epsilon$ on the experimental pure-state output density matrix $\rho_{exp}$, 
 defined as $\epsilon = \parallel\rho_{exp} - 
 \rho_{theory}\parallel/\parallel \rho_{theory}\parallel$,  varied between $8$
 and $12\%$.} 
\label{fig:denmat_pure}
\end{figure}

We also characterized the entire deviation density matrix $\rho_\Delta \equiv
\rho - {\rm Tr}(\rho) I/4$ (Fig.~\ref{fig:denmat_pure}) describing the final
2-qubit state.  These results unambiguously demonstrate the complete proper
functioning of the quantum algorithm, and provide data for the error analysis
described below.

Quantum computation requires that a coherent superposition be preserved for
the duration of the computation. This requires a highly isolated quantum
system (small $\hat{\cal H}_{env}$), and fortunately, nuclear spins are
naturally well-isolated from their environment.  Phase randomization due to
${\bf B}_0$ inhomogeneities was minimized by using about 30 electromagnetic
coils to shim the static field to be constant to about one part in $10^9$
over the sample volume. The longitudinal and transverse relaxation time
constants $T_1$ and $T_2$ were measured using standard inversion-recovery
and Carr-Purcell-Meiboom-Gill pulse sequences~\cite{Ernst94}, giving
$T_1\approx 19$ and $25$~seconds, and $T_2\approx 7$ and $0.3$ seconds,
respectively, for proton and carbon; these were much longer than required for
our experiment, which finished in about $7$ milliseconds.

The single most important source of errors in the experiments was
the RF field inhomogeneity and pulse length calibration imperfections.
A direct measure of this inhomogeneity is the $\approx 200$ $\mu$s time 
constant of the exponentially decaying envelope observed from applying a single
pulse, as a function of pulse width.  Including the population
permutation sequence, about 7 pulses are applied to each nucleus, with
a cumulative duration of $\approx 70-100 \mu s$.

The second most important contribution to errors is the low carbon
signal-to-noise ratio, $\mbox{signal peak height}/{\mbox{RMS noise}}
\approx 35$, versus $\approx 4300$ for proton. The carbon signal was
much weaker because the carbon gyromagnetic ratio is 4 times smaller,
and the carbon receiver coil is mounted more remotely from the sample.
Smaller contributions to errors came from incomplete relaxation
between subsequent experiments, carrier frequency offsets, and
numerical errors in the data analysis.

For this small-scale quantum computer, imperfections were dominated by
technology, rather than by fundamental issues. However, NMR
quantum computers larger than about $10$ qubits will require creative new
approaches, since the signal strength decays exponentially with the number of
qubits in the machine, using current schemes\cite{Chuang97e,Warren97}: for $N$
spins the signal from the initial state $00\ldots 0$ is proportional to
$n_{00\ldots 0} \propto N Z^{-N}$ where the single spin partition function
$Z\approx 2$ at high temperatures.
Furthermore, coherence times typically decrease for larger molecules, while
the average logic gate duration increases.  Nevertheless, there is hope; for
example, due to the ensemble nature of the NMR approach, one can infer the
output result as long as a {\em distinguishable majority} of the molecules
reaches the correct final state.  Creating an effective pure state is thus
not always necessary, as we have demonstrated.  Optical pumping and other
cooling techniques can also be used to pre-polarize the sample to increase
the output signal amplitude, since $Z\approx 1$ at low temperatures.
Quantum computation clearly poses an interesting and relevant experimental
challenge for the future.


Note: during the preparation of this manuscript we became aware of a closely
related experiment by J.A. Jones and M. Mosca at Oxford
University\cite{Jones98}.  \\

 


{\em Acknowledgments}\\

\footnotesize
We thank Alex Pines and Mark Kubinec for helpful discussions. This work was 
supported by DARPA under the NMRQC initiative. L.V. gratefully acknowledges a 
Francqui Fellowship of the Belgian American Educational Foundation and  a 
Yansouni Family Fellowship.\\

Correspondence and requests for materials should be addressed to I.L. Chuang,
electronic address ichuang@almaden.ibm.com\\

\end{document}